 \documentclass[pmlr,twocolumn,10pt]{jmlr} 

\usepackage{booktabs}
\usepackage{siunitx}

\usepackage[switch]{lineno}

\theorembodyfont{\upshape}
\theoremheaderfont{\scshape}
\theorempostheader{:}
\theoremsep{\newline}

\jmlrvolume{}
\jmlryear{2024}
\jmlrsubmitted{}
\jmlrpublished{}
\jmlrworkshop{Machine Learning for Health (ML4H) 2024} 

 \title[Economic Implications of Using Machine Learning in Clinical Psychiatry]{Evaluating the Economic Implications of Using Machine Learning in Clinical Psychiatry}

\author{%
\Name{Soaad Hossain} \Email{soaad.hossain@mail.utoronto.ca}\\
\addr University of Toronto, Canada
\AND
\Name{James Rasalingam} \Email{james.rasalingam22@ac.uk}\\
\addr Imperial College London, England
\AND
\Name{Arhum Waheed} \Email{arhum.waheed@mail.utoronto.ca}\\
\addr University of Toronto, Canada
\AND
\Name{Fatah Awil} \Email{fawil@hrh.ca}\\
\addr University of York, England
\AND
\Name{Rachel Kandiah} \Email{rkandiah@uwo.ca}\\
\addr Western University, Canada
\AND
\Name{Syed Ishtiaque Ahmed} \Email{ishtiaque@cs.toronto.edu}\\
\addr University of Toronto, Canada
}

\nolinenumbers 

\begin{document}

\maketitle

\begin{abstract}
With the growing interest in using AI and machine learning (ML) in medicine, there is an increasing number of literature covering the application and ethics of using AI and ML in areas of medicine such as clinical psychiatry. The problem is that there is little literature covering the economic aspects associated with using ML in clinical psychiatry. This study addresses this gap by specifically studying the economic implications of using ML in clinical psychiatry. In this paper, we evaluate the economic implications of using ML in clinical psychiatry through using three problem-oriented case studies, literature on economics, socioeconomic and medical AI, and two types of health economic evaluations. In addition, we provide details on fairness, legal, ethics and other considerations for ML in clinical psychiatry. 

\end{abstract}

\paragraph*{Data and Code Availability}
Data was not used for this research. 

\paragraph*{Institutional Review Board (IRB)}
This research does not require IRB approval.

\section{Introduction}
\label{sec:intro}
With the success of artificial intelligence (AI) and machine learning (ML) within areas such as transportation and finance, there is an increasing interest in using those within areas of medicine. One of the areas of interest is psychiatry. There is a growing interest in applying ML in clinical practice within psychiatry, and more recently, research is being conducted to understand its effectiveness in psychiatry. However, no research has investigated the economic implications associated with its use. Our study addresses this gap by evaluating the economic implications of using ML in clinical psychiatry. This paper will first review the current situation of clinical psychiatry and economics, and the economic, socioeconomic, and medical incentives for ML in clinical psychiatry. Then, using three cases studies, we will evaluate the economic implications of using ML in clinical psychiatry. We will conclude by discussing ethical, legal and other considerations for ML in clinical psychiatry.

\section{Economics and the Economic Burden of Mental Disorders}
\label{sec:state}
Economics is concerned with producing, distributing, and consuming goods and services \citep{Knapp_2020}. When it comes to economics and mental health, there are multiple types of health economic evaluations. Two of those are cost-effectiveness analysis (CEA) and cost-benefit analysis (CBA). We will use both of those for our evaluation. Results from CEAs help decision-makers by providing information on the additional cost of achieving an incremental improvement in an outcome measure \citep{Knapp_2020}. Results from CBAs, from a societal and public health perspective, are expressed in net benefits \citep{Knapp_2020}.

The best way to understand the current state of psychiatry from an economic perspective is by reviewing the scale of the mental illnesses problem and highlighting the mismatch between mental health burden and resource allocation. The World Health Organization (WHO) World Mental Health Report finds that one billion people (more than one in eight adults and adolescents) worldwide have a mental disorder \citep{Cuijpers_2023}. Mental illnesses are also financially costly as many mental illnesses affect working-age people, which in turn impacts production. It has been estimated that globally speaking, 12 billion productive workdays are lost every year to depression and anxiety alone, at the cost of nearly 1 trillion dollars in the US \citep{Cuijpers_2023}. Analyses have suggested that the global cost of mental, neurological and substance use disorders is 2.5 trillion dollars in the US and reduced productivity in 2010 \citep{Trautmann_2016}. In 2019, the global cost is estimated at 5 trillion dollars US \citep{Arias_2022}. Despite the high prevalence, enormous disease burden and massive economic costs of mental disorders, the majority of mental health services across the world fail to meet the mental health needs of their population \citep{Cuijpers_2023}.

It is important to note that from a decision-makers' perspective, evidence on disease burden and costs alone has limited use as it does not offer recommendations on solutions (e.g., treatments, services, or preventative measures) \citep{Knapp_2020}. However, economic evaluation studies that consider both cost and outcomes can help them make decisions that consider affordability and  the cost-benefit.

\section{Economic Incentives for Machine Learning in Clinical Psychiatry}
\label{sec:econincentives}
AI has proved in manufacturing and other sectors that it can increase the efficiency of processes, improve user experience, reduce information asymmetry and develop new models and products \citep{Borrellas_2021}. Incremental gains from businesses through AI lead to social benefits where consumers obtain better access to services, limited resources are used more effectively, and as ML solutions scale, they become less costly and more accessible \citep{Borrellas_2021}. The success of ML-based systems has led to  growing interest in its use within healthcare and medicine, including clinical psychiatry.

Several studies have looked at the use of AI and ML in healthcare and medicine. From those studies, ML has shown tremendous promise in improving multiple areas of healthcare and medicine, such as diagnostic accuracy and disease management in African healthcare settings \citep{Arefin_2024}. One of those studies is the seminal study by Ogbaga et al., conducted in  Nigeria. Ogbaga's study demonstrated the effectiveness of  AI-powered diagnostic tools in detecting diseases with high precision \citep{Arefin_2024}. Other studies that study ML for its ability to analyze, classify, diagnose, manage, monitor, and predict different health conditions or diseases such as cancer, tuberculosis, diabetes, hypertension, mental health conditions and heart disease are listed and studied in Okeibunor et al.'s scoping review on the use of AI for delivery of essential health services across World Health Organization (WHO) regions \citep{Chukwudi_2023}. The scoping review found that most of the studies focused on using ML as an application to detect and diagnose different diseases. AI  and ML solutions enable more successful diagnoses and hasten the diagnostic process, ultimately giving healthcare providers more time to perform other tasks and see more patients, reducing costs for institutions. The costs are further reduced as misdiagnoses and the number of readmission due to incorrect diagnoses are reduced \citep{Ericson_2024}.

The economic incentives for ML in healthcare and medicine relevant to clinical medicine are essentially the same as those for ML in clinical psychiatry. Ultimately, ML can assist psychiatrists, physicians, researchers, psychologists, and other mental health professionals in completing the clinical work needed for screening \citep{Wright-Berryman_2023}, diagnoses, and treatments \citep{Bzdok_2018}, reducing the costs associated with those tasks.

It is important to note that while many studies illustrate how AI and ML can reduce costs and resources, some studies highlight the substantial requirements needed for ML solutions in healthcare and medicine, which should be considered when investigating the economic aspects of ML for healthcare and medicine. The studies highlight how implementing and scaling ML algorithms within healthcare systems requires substantial resources, infrastructure, expertise, and adequate endorsement at the clinical end-user, departmental, and institutional levels \citep{Kwong_2024}. They also mention how such an ecosystem may be challenging outside of academic settings or within single-payer healthcare systems, and how the costs and benefits of ML should be carefully considered through health technology assessments because their incremental advantages may not always justify the steep costs required to implement and maintain such solutions \citep{Kwong_2024}.

\section{Socioeconomic Incentives for Machine Learning in Clinical Psychiatry}
\label{sec:socioeconincentives}
The socioeconomic incentives for machine learning in clinical psychiatry primarily stem from facts on low socioeconomic status and mental illnesses. There is a clear link that exists between social and economic inequality and poor mental health, and there is growing evidence that supports this \citep{Macintyre_2018}. Experience of health equity issues and socioeconomic disadvantage, including low income, poverty and poor housing, is consistently associated with poorer mental health. Greater socioeconomic inequality within societies is associated with a greater prevalence of mental illness \citep{Macintyre_2018}.

Inflation is an economic determinant of psychological well-being \citep{Prati_2023}. With inflation, issues accessing resources and other social, economic and financial problems increasing around the world, the number of people experiencing varying degrees of socioeconomic disadvantage is increasing, creating a growing demand for mental health services across all regions and areas around the world (\cite{Addison_2019}; \cite{Chancel_2022}). This demand, however, is a significant problem for healthcare, medicine and society as the supply of mental health professionals is very limited, posing a problem for those trying to access mental health services. Realistically speaking, there are only three solutions that could tackle the growing demand for mental health services. The first is increasing the number of mental health professionals. The second is empowering existing mental health professionals to serve more people in less time. The third is the combination of the first and second solutions. The first is difficult to achieve due to the time and resources needed to educate and train psychiatrists and other mental health professionals. As a result, the third solution is challenging, leaving the second one as the only viable option. 

ML can be used as a tool to achieve the second option. It can help mental health professionals with their clinical tasks, enabling them to complete their work quicker and see more patients with fewer resources. In turn, this allows them and institutions to serve more individuals of low, medium, and high socioeconomic status in less time and with the resources they have. These solutions are essential for people experiencing health equity problems and socioeconomic disadvantages as this can help prevent their situation and condition from worsening \citep{Ericson_2024}.

\section{Medical Incentives for Machine Learning in Clinical Psychiatry}
\label{sec:medincentives}
When considering medical incentives for ML in clinical psychiatry, we must consider two incentives: supervised and unsupervised learning. Note that other types of incentives can and should be considered, such as semi-supervised learning \citep{Chen_2022}; however, for the scope of this paper, we will only cover supervised and unsupervised learning.

The medical incentives for supervised learning in clinical psychiatry are primarily based on using supervised learning to predict psychiatric disorders and help with tasks associated with predicting psychiatric disorders (i.e., diagnosis). Systematic reviews such as that conducted by Pgoni et al. found that most machine learning algorithms used for predicting suicide were random forests (RF) and support vector machine (SVM) \citep{Pigoni_2024}. However, the use of supervised learning is not limited to suicide prediction. It extends to other mental health issues, and it has other ways of supporting the process of diagnosing mental illnesses. For example, relevance vector machine (RVM) has recently demonstrated its ability to quantify neuroimaging biomarkers for post-traumatic stress disorder (PTSD) diagnosis \citep{Chen_2022}. Similarly, a study using decision trees (DT) found that DT can effectively score and interpret psychodiagnostic test data and generate flowcharts that can guide psychiatrists in the diagnostic assessment process \citep{Colledani_2023}. 

The secondary incentives for supervised learning in psychiatry are using supervised learning as a screening tool and assisting with psychiatric diagnoses. Some studies support the idea of using supervised learning for screening patients. One of those studies is the study by Wright-Berrman et al. In their study, they found that it is feasible to use a virtual platform that uses logistic regression (LR), SVM and extreme gradient boosting (XGB) to screen for depression, anxiety and suicide risk \citep{Wright-Berryman_2023}. SVMs have also shown that they could be used to understand treatments. Past studies have investigated the use of SVM to predict response to ketamine treatment in major depressive disorder patients and found that ML methods such as SVM could significantly improve the prediction of treatment efficacy, offering potential benefits for personalized treatments and healthcare cost reduction \citep{Romão_2024}.  

The medical incentives for unsupervised learning in clinical psychiatry are primarily diagnosis and secondary treatment for mental disorders. Unsupervised learning has demonstrated that it could be used to help clinicians with diagnosing mental illnesses such as Alzheimer's \citep{Liu_2024}, PTSD \citep{Wang_2024}, eating disorders \citep{Ghosh_2024}, and autism spectrum disorder \citep{Koehler_2024}. Unsupervised learning can also assist with the treatment of mental illnesses. For instance, in a study by Garcia-Rudolph et al., they found that clustering could be used to group individuals with acquired brain injury into different subgroups based on their neuropsychological profiles, and web-based cognitive rehabilitation can then be utilized and tailored to the specific needs of each subgroup \citep{Khorev_2024}. In both the case of diagnosis and treatment, unsupervised learning could help mental health professionals with those.

\section{Selecting Psychiatric Use Cases for Evaluation}
\label{sec:selection}
We find sufficient motivation for a handful of psychiatric use cases using the literature on economic, socioeconomic and medical incentives for ML in clinical psychiatry. However, to stay within the scope of the paper, we will only focus on selected use cases within the handful. Those use cases are the following: (1) Using ML as a screening tool for mental health issues; (2) Using ML as an assistive tool for mental health professionals; and (3) Using ML to replace mental health professionals in low mental health professionals supply-high demand areas. Each case will be described to illustrate the case from a healthcare and medical perspective. An evaluation from a health economic or economic perspective will then follow the description.

\section{Evaluation Process}
\label{sec:process}
For the first and second case study, we will use CEA and CBA respectively. Accordingly, the evaluation will focus on understanding cost minimization, utility and effectiveness. Then, the findings from the evaluation will be used to extract economic implications. There are some critical notes to keep in mind for this evaluation. In addition, all of our cases primarily focuses on adult patients.

Since ML-based systems have yet to be implemented within hospitals and other healthcare settings, it is not currently possible to precisely and numerically calculate health economic evaluations using the formulas for CEA, among others. The next step, at a much later time, would be to perform these calculations and publish the results in studies. Such studies would be economic evaluation studies. Our study is a problem-oriented case study to help understand economic implications through solving problems in clinical psychiatry. This will help decision-makers understand the economic implications behind using ML in clinical psychiatry and academics and others to better approach those economic evaluation studies. This will also help healthcare and medical administrators and professionals better understand the use of AI and ML in clinical medicine. 

For our evaluation, we will use a problem-oriented case study approach that uses health economic evaluations, an understanding of the technical and economic aspects of AI and ML in healthcare and medicine, and variables involved in psychiatric cases to evaluate and detail the economic implications.

\section{Evaluating Economic Implications Using Selected Use Cases}
\label{sec:evaluation}

\subsection{Case 1: Machine Learning as a Screening Tool in Clinical Psychiatry}
The case considering ML as a screening tool in clinical psychiatry would entail using an ML-based application that utilizes a interview or questionnaires via websites and mobile phones to assess respondents' emotional and psychological states rapidly and access to early detection of psychological concerns such as symptoms of anxiety, depression, stress, and other moderate or severe manifestations of mental disorders \citep{Chamorro-Delmo_2024}. This innovative screening approach provides a swift and personalized evaluation of those individuals who may require intervention, as it facilitates access to available healthcare resources \citep{Chamorro-Delmo_2024}. For our case, we will use the health economic evaluation CEA as CEA allows for a simple cost-and-effect comparison of alternative interventions \citep{Lau_2017}. 

We will refer to the ML-based screening tool as a smart screening tool. The two interventions that we will compare are the smart screening tool and the current screening tools used in clinical psychiatry, specifically digital mental health screening tools as those are said to provide numerous benefits such as minimal delivery costs, efficient data collection, and increased convenience \citep{Martin_2022}.

Current digital (and non-digital) screening tools in clinical psychiatry are screening questionnaires used by psychiatrists and other mental health professionals to see whether or not a person may have a mental health issue \citep{Shields_2021}. The tool uses one of the following screening questionnaires: Structured Inventory of Malingered Symptomatology (SIMS), Pain Patient Profile (P-3), or Personality Assessment Inventory (PAI). SIMS is a 75-item, true-or-false screening assessment that assesses malingered psychopathology and neuropsychological symptoms in adults \citep{par_inc}. P-3 is an assessment with 44 items that focuses on the factors most often associated with chronic pain and can help provide an objective link between the physician's observations and the possible need for further psychological assessment \citep{pearson}. PAI is a 344-item assessment that provides information relevant to clinical diagnosis, treatment planning and screening for psychopathology and covers constructs most relevant to a broad-based assessment of mental disorders \citep{sigma_inc}. Given the cost, time, and effort needed for PAI and the SIMS use case, P-3 is the questionnaire often used in digital mental health screening tools. Note that most digital mental health tools were designed to target a single condition rather than comprehensive assessments of psychopathology, with most including $<$45 questions; a combination of these tools is recommended for comprehensive mental health assessments \citep{Martin_2022}. After the screening assessment is completed, the digital mental health tool generates a report with the results from the assessment and insight into what mental health issue the patient may have. The mental health professional uses the report and the results to score the assessment. The next step in the clinical psychiatric process occurs if the score is valid. If the score is invalid, the mental health professional will conduct a follow-up interview with the patient for further details to help them decide. There are two types of costs to consider. The first is the cost of using a single digital mental health screening tool and the cost of work from the mental health professional. The second type is the cost associated with using two or more digital mental health screening tools and the cost of the work from the mental health professional.  

With a smart screening tool, there are two parts to them. The first part is the screening assessment component. The questions can be asked through a questionnaire similar to digital mental health tools or an interview. The second part, done after the screening questions are complete, is the detection of potential mental disorders based on the answers. Unlike current digital screening mental health tools, only one smart screening tool is needed for simple and comprehensive mental health assessments \citep{Chamorro-Delmo_2024}. The reason for this is how ML algorithms are designed. ML algorithms are designed to incorporate a large amount of data and use it for classification, regression, clustering, anomaly detection, ranking, recommendation, and forecasting purposes. For example, Principal Component Analysis (PCA) in ML can be used as an anomaly detection model by the smart screening tool to reveal the inner structure of the patient's data and explain the variance in that data. This information and information obtained through another ML model within the smart screening tool (e.g. SVM) can provide explanations, diagnoses and recommendations. After both parts are complete, the smart screening tool generates a report with the assessment results, a recommended decision on the patient, and a diagnosis of the mental disorder based on those results (in the case where the tool found that the patient has a mental disorder). The mental health professional then uses the report and recommendation to decide whether to move on to the next step in the clinical psychiatric process or dismiss the patient. In rare cases, the mental health professional will conduct a follow-up interview with the patient for further details to help them decide. The cost associated with using the smart screening tool is the cost of the tool and the cost of the work done by mental health professionals.

Using CEA, we can compare the cost and effect of a single digital mental health screening tool versus a single smart screening tool, and the cost and effect of multiple digital mental health screening tools versus a single smart screening tool. With CEA, costs are assessed by identifying the study's perspective and all the resources used and quantifying them into physical units. With a single digital mental health screening tool, the risk of misdiagnosis is higher. In addition, using a single digital mental health tool leaves more work for the mental health professional as it does not provide sufficient information for them to make a diagnosis efficiently. This results in more cost and effort from them. In contrast, a smart screening tool is more accurately able to diagnose the mental health issue in the patient and provide a more detailed report to support diagnosis, making the diagnosis process easier for mental health professionals and resulting in less cost and effort.

With multiple digital mental health screening tools, the risk of misdiagnosis is lower, while the overall cost will be higher as multiple screening tools are used. Multiple resources are used, and the exact quantity of physical units depends on the number of digital mental health screening tools used. As multiple digital mental health tools are used, the cost and effort needed increase. In addition, there is still scoring and other work that mental health professional needs to do, increasing the cost associated with their use. The smart screening tool could be seen as a single digital mental health screening tool cost-wise. However, it is better as, unlike digital mental health screening tools, it can incorporate a much larger set of factors and conditions, providing better reporting, predictions, and recommendations. As a result, the work and effort needed from a mental health professional are minimized. The work has a cost, but it is less than using one or more digital mental health screening tools, as the mental health professional does not need to perform scoring and other tasks. For both the case of single and multiple digital mental health screening tools, the simple CEA equation used to compare the cost and effect of them with the smart screening tool is the following:

\begin{equation} \label{eq:C}
C(X) = A(X) - B(X) 
\end{equation}

Where

\begin{align}
    A(X) = X((S(X) + D(X)) + P(X)) + \varepsilon \label{eq:A} \\
    B(X) = M(X) + L(X) + \frac{1}{3}P(X) + \varepsilon \label{eq:B}
\end{align}

\equationref{eq:C} is the total cost and effort difference between using the digital mental health screening tool(s) and the smart screening tool. \equationref{eq:A} is the total cost of and using one or more digital mental health screening tool(s) and work from the mental health professional. \equationref{eq:B} is the total cost of and using the smart screening tool and work from the mental health professional. $X$ is the number of digital mental health screening tools used, $S(X)$ is the cost of each digital mental health screening tool, $D(X)$ is the cost of using each digital mental health screening tool, $P(X)$ is the cost of the work done by the mental health professional, $M(X)$ is the cost of a smart tool, $L(X)$ is the cost of using the smart tool. $\varepsilon$ is the cost of medical errors in the screening process (e.g. misdiagnosis). The 1/3 comes from the fact that the mental health professional does not need to do scoring and further assessments or interviews. The full CEA equation would require additional variables such as time. However, the simple CEA equation, \equationref{eq:C}, is enough to understand the cost and effect comparison between the tools. It is important to note that even if somehow cost and effect comparison X((S(X) + D(X)) in \equationref{eq:A} is equal to M(X) + L(X) in \equationref{eq:B}, \equationref{eq:A} will always be greater than \equationref{eq:B} due to P(X), the work (and workload) of the mental health professional. This implies that institutions and other entities using smart screening tools will save more on costs than with one or more digital mental health screening tools.  

\subsection{Case 2: Machine Learning Solutions as an Assistive Tool in Clinical Psychiatry}
The most realistic use of ML in clinical decision-making is using ML as an assistive tool for mental health professionals. It entails mental health professionals using the ML solutions to assist them with diagnoses and treatments of mental disorders. We will evaluate this case for economic implications using the economic evaluations CBA.

The ideal way to perform the CBA evaluation would be by using and comparing the costs associated with providing the mental health service along with the costs associated with the time needed by psychiatry and other mental health professionals to deal with psychiatric cases with and without ML to determine if it is more beneficial to use ML solutions as an assistive tool in clinical psychiatry. However, since the use of AI and ML has not been deployed in clinical psychiatric settings, performing such evaluation effectively is currently impossible. However, we will use our understanding of economics, ML, and clinical psychiatry to perform an alternative type of CBA evaluation. 

For the CBA evaluation, three types of benefits need to be considered: immediate, short-term, and long-term. In terms of immediate benefits, those benefits could be understood through the existing literature on medical AI. The main takeaway from the literature is that ML can help mental health professionals with performing screenings, diagnoses and treatments of mental disorders (\cite{Dwyer_2018}; \cite{Chekroud_2021}). Accordingly, the economic implication is that ML would reduce the costs for mental health professionals, institutions, and other entities. ML would reduce the cost associated with performing screenings, diagnoses and treatments by reducing the time and effort needed to complete those tasks with less time and resources. In turn, this reduces costs for institutions and others, enabling mental health professionals to attend to more patients in a fixed amount of time.  

The short-term benefits follow the immediate benefits. In the short, and arguably long term, using ML in clinical psychiatry improves the performance of psychiatrists and other mental health professionals. In ML helping perform screenings, diagnoses and treatments of mental disorders, it reduces the cognition and mental effort needed from mental health professionals for those tasks, enabling mental health professionals to perform effectively for more extended periods. As a result, costs for institutions and other entities are reduced as medical errors (due to non-material reasons such as the use of broken measurement tools) are reduced, and the resources needed for mental disorder screening, diagnosing and treating psychiatric patients are reduced \citep{Yan_2023}. 

The long-term benefits mostly stems from the aftermath of immediate and short-term benefits. ML helping mental health professionals serve more people in less time allows them, institutions and other entities to tackle the growing demand for mental health services by allowing them to take on and serve more people in less time, reducing costs associated with resources. The improvement in the performance of mental health professionals through using ML does provide long-term benefits cost-wise as the improvement helps reduce the number of patients readmitted (due to misdiagnosis, or wrong treatment), the number of medical errors and costs associated with medical errors, and increasing efficiencies in mental disorder screening, diagnosis and treatment process.

With all the types of benefits, we obtain the simple CBA equation used to compare the cost concerning the benefits of using ML solutions as an assistive tool for clinical psychiatry, which is the following:

\begin{equation} \label{eq:N}
N(X) = T(X) - U(X) 
\end{equation}

Where 

\begin{align}
    U(X) = \int_{0}^{t} (H_{t}(X) - \varepsilon) \,dx \label{eq:U} \\
    T(X) = \int_{0}^{t} (E_{t}(X) - \varepsilon) \,dx \label{eq:T}
\end{align}

\equationref{eq:N} is the difference in net benefit from choosing to use ML solutions as an assistive tool for clinical psychiatry. \equationref{eq:U} is the net benefits of using solely mental health professionals over some time (immediate, short-term, long-term). \equationref{eq:T} is the total benefits of using an ML solution as an assistive tool over some time (immediate, short-term, long-term). $\varepsilon$ is the cost of medical errors in the diagnosis and treatment (e.g. misdiagnosis). The full CBA equation would require additional variables such as the cost associated with purchasing and deploying an ML-based system (which that variable would be included in the equation $E_{t}(X)$. However, the simple CBA equation, \equationref{eq:N}, is enough to understand the difference in benefits between using and not using ML-based systems as a tool for clinical psychiatry. It is important to note that ML is continuously being worked on to improve its ability to complete classification and other tasks needed for diagnosis and treatments. For example, optimization approaches such as precision-focused optimization techniques are being worked on. Precision-focused optimization techniques such as Stochastic Gradient Descent (SGD), Adam, AdaGrad, RMSprop, Nesterov Accelerated Gradient (NAG), and Adadelta are crucial for refining ML models to get superior prediction accuracy and reducing error measures \citep{Bian_2024}. As ML is optimized, the net benefits that it provides for mental health professionals will further increase. As a result, the net benefit of using ML as an assistive tool for clinical psychiatry will always be higher than the net benefit of only using mental health professionals.

\subsection{Case 3: Machine Learning Solutions Replacing Mental Health Professionals in Low Supply-High Demand Areas}
The case of ML solutions replacing mental health professionals is a case that can, and should, be considered in areas where there is no or inadequate supply of mental health professionals but a high demand for mental health services. These areas are generally rural areas where healthcare resources are scarce. Such areas include areas with a high concentration of indigenous people and marginalized communities, such as Nunavut, Canada \citep{govca_2023}. In this case, the ML-based systems are performing mental disorder screening, diagnosis and treatment for people seeking mental health services in rural and other areas with little to no mental health services offered by mental health professionals.

The economic implications of using ML in such a manner differ from the economic implications of the other cases, as the focus is primarily on how using ML-based systems in such settings can indirectly reduce costs for institutions and other entities. ML solutions reduce costs by reducing and preventing incidents and cases such as readmission. The problem with rural and other areas with a low supply of mental health services is that there are people in those areas who suffer from mental illnesses who are not receiving treatment for it, consequently causing them to develop other problems such as addiction to alcohol or other substances (\cite{Friesen_2022}; \cite{Kirchebnera_2022}). Subsequently, those people or people impacted by them (e.g. victims of drunk driving accidents, domestic violence, and various forms of assaults) are eventually admitted to hospitals and other healthcare facilities for treatment, costing hospitals and other institutions resources \citep{Kirchebnera_2022}. These costs are a significant problem for them as healthcare resources in such areas are minimal, and using resources on such preventable cases prevents them from using resources on unpreventable cases (e.g., hereditary diseases/genetic disorders) and accidental cases (e.g., elderly accidentally falling and breaking a bone as a result). ML solutions taking over the role of mental health professionals can fill in the gap in mental health services in those regions, allowing those people suffering from mental illnesses to receive the treatment needed for them to function correctly. Receiving treatment can reduce preventable harmful cases, reducing the number of hospital admissions and resources needed by hospitals and other institutions as it enables existing resources to attend to other patients and cases while also saving patients' lives \citep{Ericson_2024}.

\section{Fairness, Legal, Ethics and Other Considerations for Machine Learning in Clinical Psychiatry}
\label{sec:considerations}
Two highly important aspects that were incorporated but not elaborated on within the case evaluations are the upfront, implementation, maintenance and change management cost of ML,and the current technical and ethical issues with ML. For the former, in the CEA and CBA equations, we incorporated those in the total cost equations (e.g., \equationref{eq:B}). Regarding the latter, it is known that ML suffers from problems such as biases, which can lead to issues like misdiagnosing a person and exacerbating existing issues \citep{Fiske_2019}. Those problems can lead to an increase in hospital admissions, and consequently an increase in the number of resources needed within hospitals and other entities. In the CEA and CBA equations, we incorporated those problems as medical errors $\varepsilon$ as those problems would be categorized as medical errors (i.e. malfunction of equipment). However, it is important to note that the cost of such errors can be high to the point where it becomes less cost effective and beneficial to use ML in those clinical psychiatric cases.

Another problem (and cost) that was not included in the evaluation is the legal aspects associated with using ML in clinical psychiatry. In the case where ML is found to cause significant damage, legal action would ensue. This would result in an additional cost for mental health professional, hospital and other entities involved that can not only remove any cost and benefits from using ML, but introduce dept as well. These risks should carefully be considered when considering the costs associated within using ML.  

\section{Conclusion}
\label{sec:conclusion}
ML solutions can help mental health professionals, hospitals and other entities with completing screening, diagnosis, treatment and other clinical tasks, improving performance, reducing medical errors, serve more patients in less time and with less resources, reduce and prevent readmission, and increasing efficiencies. In turn, these would reduce costs for mental health professionals, hospitals and other entities. In addition, the upfront, implementation, maintenance and change management cost of ML solutions, medical errors from using ML and potential legal costs with using ML should be considered when considering the cost of using ML for clinical psychiatry. 

\acks{Acknowledgments go here \emph{but should only appear in the
camera-ready version of the paper if it is accepted}.
Acknowledgments do not count toward the paper page limit.}

\bibliography{jmlr-sample}

\begin{thebibliography}{39}
\providecommand{\natexlab}[1]{#1}
\providecommand{\url}[1]{\texttt{#1}}
\expandafter\ifx\csname urlstyle\endcsname\relax
  \providecommand{\doi}[1]{doi: #1}\else
  \providecommand{\doi}{doi: \begingroup \urlstyle{rm}\Url}\fi

\bibitem[Addison et~al.(2019)Addison, Pirttilä, and Tarp]{Addison_2019}
T.~Addison, J.~Pirttilä, and F.~Tarp.
\newblock Is global inequality rising or falling?
\newblock \emph{United Nations University World Institute for Development Economics Research (UNU-WIDER)}, 2, February 2019.

\bibitem[Arefin(2024)]{Arefin_2024}
S.~Arefin.
\newblock Leveraging ai for healthcare advancement in africa.
\newblock \emph{Academic Journal of Science and Technology}, 7:\penalty0 1--11, January 2024.

\bibitem[Arias et~al.(2022)Arias, Saxena, and Verguet]{Arias_2022}
D.~Arias, S.~Saxena, and S.~Verguet.
\newblock Quantifying the global burden of mental disorders and their economic value.
\newblock \emph{eClinicalMedicine}, 54:\penalty0 101675, December 2022.

\bibitem[Bian and Priyadarshi(2024)]{Bian_2024}
K.~Bian and R.~Priyadarshi.
\newblock Machine learning optimization techniques: A survey, classification, challenges, and future research issues.
\newblock \emph{Archives of Computational Methods in Engineering}, March 2024.

\bibitem[Borrellas and Unceta(2021)]{Borrellas_2021}
P.~Borrellas and I.~Unceta.
\newblock The challenges of machine learning and their economic implications.
\newblock \emph{Entropy}, 23:\penalty0 275, December 2021.

\bibitem[Bzdok and Meyer-Lindenberg(2018)]{Bzdok_2018}
D.~Bzdok and A.~Meyer-Lindenberg.
\newblock Machine learning for precision psychiatry: Opportunities and challenges.
\newblock \emph{Biological Psychiatry: Cognitive Neuroscience and Neuroimaging}, 3:\penalty0 223--230, March 2018.

\bibitem[Chamorro-Delmo et~al.(2024)Chamorro-Delmo, Lopez-Fernandez, Villasante-Soriano, Antonio, Álvarez García, Porras-Segovia, and Baca-García]{Chamorro-Delmo_2024}
J.~Chamorro-Delmo, O.~Lopez-Fernandez, P.~Villasante-Soriano, P.~Portillo-De Antonio, R.~Álvarez García, A.~Porras-Segovia, and E.~Baca-García.
\newblock A feasibility study of a smart screening tool for people at risk of mental health issues: Response rate, and sociodemographic and clinical factors.
\newblock \emph{Journal of Affective Disorders}, 362:\penalty0 755--761, October 2024.

\bibitem[Chancel et~al.(2022)Chancel, Piketty, Saez, and et~al.]{Chancel_2022}
L.~Chancel, T.~Piketty, E.~Saez, and G.~Zucman et~al.
\newblock World inequality report 2022.
\newblock \emph{World Inequality Lab}, February 2022.

\bibitem[Chekroud et~al.(2021)Chekroud, Bondar, Delgadillo, Doherty, Wasil, Fokkema, Cohen, Belgrave, DeRubeis, Iniesta, and Choi]{Chekroud_2021}
A.~M. Chekroud, J.~Bondar, J.~Delgadillo, G.~Doherty, A.~Wasil, M.~Fokkema, Z.~Cohen, D.~Belgrave, R.~DeRubeis, R.~Iniesta, and D.~Dwyerand~K. Choi.
\newblock The promise of machine learning in predicting treatment outcomes in psychiatry.
\newblock \emph{World Psychiatry}, 20:\penalty0 154--170, May 2021.

\bibitem[Chen et~al.(2022)Chen, Kulkarni, Galatzer-Levy, Bigio, Nasca, and Zhang]{Chen_2022}
Z.~S. Chen, P.~P. Kulkarni, I.~R. Galatzer-Levy, B.~Bigio, C.~Nasca, and Y.~Zhang.
\newblock Modern views of machine learning for precision psychiatry.
\newblock \emph{Patterns}, 3:\penalty0 100602, November 2022.

\bibitem[Chukwudi et~al.(2023)Chukwudi, Anelisa, Juliana, Ngozi, Housseynou, Pamela, Mavis, Edison, Derrick, Shey, and Lindiwe]{Chukwudi_2023}
O.~J. Chukwudi, J.~Anelisa, I.~C. Juliana, I.~Ngozi, B.~Housseynou, Z.~Zukiswa Pamela, N.~A. Mavis, M.~Edison, M.~Derrick, W.~C. Shey, and M.~Lindiwe.
\newblock The use of artificial intelligence for delivery of essential health services across who regions: a scoping review.
\newblock \emph{Frontiers in Public Health}, 11:\penalty0 1102185, July 2023.

\bibitem[Colledani et~al.(2023)Colledani, Anselmi, and Robusto]{Colledani_2023}
D.~Colledani, P.~Anselmi, and E.~Robusto.
\newblock Machine learning-decision tree classifiers in psychiatric assessment: An application to the diagnosis of major depressive disorder.
\newblock \emph{Psychiatry Research}, 322:\penalty0 115127, April 2023.

\bibitem[Cuijpers et~al.(2023)Cuijpers, Javed, and Bhui]{Cuijpers_2023}
P.~Cuijpers, A.~Javed, and K.~Bhui.
\newblock The who world mental health report: a call for action.
\newblock \emph{The British Journal of Psychiatry}, 222:\penalty0 227–229, June 2023.

\bibitem[Dwyer et~al.(2018)Dwyer, Falkai, and Koutsouleris]{Dwyer_2018}
D.~B Dwyer, P.~Falkai, and N.~Koutsouleris.
\newblock Machine learning approaches for clinical psychology and psychiatry.
\newblock \emph{Journal of Health Economics and Outcomes Research}, 7:\penalty0 91--118, January 2018.

\bibitem[Ericson et~al.(2024)Ericson, Hjelmgren, Sjövall, Söderberg, and Persson]{Ericson_2024}
O.~Ericson, J.~Hjelmgren, F.~Sjövall, J.~Söderberg, and I.~Persson.
\newblock The potential cost and cost-effectiveness impact of using a machine learning algorithm for early detection of sepsis in intensive care units in sweden.
\newblock \emph{Journal of Health Economics and Outcomes Research}, 9:\penalty0 101–110, April 2024.

\bibitem[Fiske et~al.(2019)Fiske, Henningsen, and Buyx]{Fiske_2019}
A.~Fiske, P.~Henningsen, and A.~Buyx.
\newblock Your robot therapist will see you now: Ethical implications of embodied artificial intelligence in psychiatry, psychology, and psychotherapy.
\newblock \emph{Journal of Medical Internet Research}, 21:\penalty0 5, May 2019.

\bibitem[Friesen et~al.(2022)Friesen, Myran, Yu, Rosella, Selby, and Kurdyak]{Friesen_2022}
E.~L. Friesen, D.~Myran, W.e Yu, L.~Rosella, P.~Selby, and P.~Kurdyak.
\newblock Rural-urban disparities in post-discharge outcomes following alcohol-related hospitalizations in ontario, canada: A retrospective cohort study.
\newblock \emph{Drug and Alcohol Dependence}, 238:\penalty0 109568, September 2022.

\bibitem[Ghosh et~al.(2024)Ghosh, Burger, Simeunovic-Ostojic, Maas, and Petković]{Ghosh_2024}
S.~Ghosh, P.~Burger, M.~Simeunovic-Ostojic, J.~Maas, and M.~Petković.
\newblock Review of machine learning solutions for eating disorders.
\newblock \emph{International Journal of Medical Informatics}, 189:\penalty0 105526, September 2024.

\bibitem[Khorev et~al.(2024)Khorev, Kiselev, Badarin, Antipov, Drapkina, Kurkin, and Hramov]{Khorev_2024}
V.~Khorev, A.~Kiselev, A.~Badarin, V.~Antipov, O.~Drapkina, S.~Kurkin, and A.~Hramov.
\newblock Review on the use of ai-based methods and tools for treating mental conditions and mental rehabilitation.
\newblock \emph{The European Physical Journal Special Topics}, August 2024.

\bibitem[Kirchebnera et~al.(2022)Kirchebnera, Laua, Habermeyera, and Sonnwebera]{Kirchebnera_2022}
J.~Kirchebnera, S.~Laua, E.~Habermeyera, and M.~Sonnwebera.
\newblock A collection of medical findings using machine learning and their relevance to psychiatry.
\newblock \emph{Swiss Archives of Neurology, Psychiatry and Psychotherapy}, 173:\penalty0 w10079, July 2022.

\bibitem[Knapp and Wong(2020)]{Knapp_2020}
M.~Knapp and G.~Wong.
\newblock Economics and mental health: the current scenario.
\newblock \emph{World Psychiatry}, 3:\penalty0 3--14, January 2020.

\bibitem[Koehler et~al.(2024)Koehler, Dong, Bierlich, Fischer, Späth, Plank, Koutsouleris, and Falter-Wagner]{Koehler_2024}
J.~C. Koehler, M.~S. Dong, A.~M. Bierlich, S.~Fischer, J.~Späth, I.~S. Plank, N.~Koutsouleris, and C.~M. Falter-Wagner.
\newblock Machine learning classification of autism spectrum disorder based on reciprocity in naturalistic social interactions.
\newblock \emph{Translational Psychiatry}, 14:\penalty0 76, February 2024.

\bibitem[Kwong et~al.(2024)Kwong, Nickel, Wang, and Kvedar]{Kwong_2024}
J.~C.~C. Kwong, G.~C. Nickel, S.~C.~Y. Wang, and J.~C. Kvedar.
\newblock Integrating artificial intelligence into healthcare systems: more than just the algorithm.
\newblock \emph{npj Digital Medicine}, 7:\penalty0 52, March 2024.

\bibitem[Langley and Tollison(2024)]{pearson}
J.~C. Langley and C.~D. Tollison.
\newblock Pain patient profile p-3.
\newblock Technical report, Pearson, 2024.

\bibitem[Lau(2017)]{Lau_2017}
F.~Lau.
\newblock ehealth economic evaluation framework.
\newblock \emph{Handbook of eHealth Evaluation: An Evidence-based Approach}, pages 93--108, February 2017.

\bibitem[Macintyre et~al.(2018)Macintyre, Ferris, Gonçalves, and Quinn]{Macintyre_2018}
A.~Macintyre, D.~Ferris, B.~Gonçalves, and N.~Quinn.
\newblock What has economics got to do with it? the impact of socioeconomic factors on mental health and the case for collective action.
\newblock \emph{Palgrave Communications}, 10:\penalty0 10, January 2018.

\bibitem[Martin-Key et~al.(2022)Martin-Key, Spadaro, Funnell, Barker, Schei, Tomasik, and Bahn]{Martin_2022}
N.~A Martin-Key, B.~Spadaro, E.~Funnell, E.~Jane Barker, T.~S. Schei, J.~Tomasik, and S.~Bahn.
\newblock The current state and validity of digital assessment tools for psychiatry: Systematic review.
\newblock \emph{JMIR Mental Health}, 9:\penalty0 3, March 2022.

\bibitem[Morey(2024)]{sigma_inc}
L.~C. Morey.
\newblock Pai personality assessment inventory.
\newblock Technical report, SIGMA Assessment Systems Inc., 2024.

\bibitem[of~Canada(2023)]{govca_2023}
Government of~Canada.
\newblock An update on the socio-economic gaps between indigenous peoples and the non-indigenous population in canada: Highlights from the 2021 census.
\newblock \emph{Annual Report to Parliament 2023}, October 2023.

\bibitem[Pigoni et~al.(2024)Pigoni, Delvecchio, N.Turtulici, Madonna, Pietrini, Cecchetti, and Brambilla]{Pigoni_2024}
A.~Pigoni, G.~Delvecchio, N.Turtulici, D.~Madonna, P.~Pietrini, L.~Cecchetti, and P.~Brambilla.
\newblock Machine learning and the prediction of suicide in psychiatric populations: a systematic review.
\newblock \emph{Translational Psychiatry}, 14:\penalty0 140, March 2024.

\bibitem[Prati(2023)]{Prati_2023}
Alberto Prati.
\newblock The well-being cost of inflation inequalities.
\newblock \emph{The Review of Income and Wealth}, 70:\penalty0 213--238, January 2023.

\bibitem[Romão et~al.(2024)Romão, Melo, R.André, and Novais]{Romão_2024}
J.~Romão, A.~Melo, R.André, and F.~Novais.
\newblock Machine learning as a tool to find new pharmacological targets in mood disorders: A systematic review.
\newblock \emph{Current Treatment Options in Psychiatry}, 11:\penalty0 241–264, August 2024.

\bibitem[Shields et~al.(2021)Shields, Korol, Carleton, McElheran, Stelnicki, Groll, , and Anderson]{Shields_2021}
R.~E. Shields, S.~Korol, R.~Nicholas Carleton, M.~McElheran, A.M. Stelnicki, D.~Groll, , and G.~S. Anderson.
\newblock Brief mental health disorder screening questionnaires and use with public safety personnel: A review.
\newblock \emph{International Journal of Environmental Research and Public Health}, 18:\penalty0 3743, April 2021.

\bibitem[Smith and Widows(2024)]{par_inc}
G.~P. Smith and M.~R. Widows.
\newblock Sims structured inventory of malingered symptomatology.
\newblock Technical report, PAR, Inc., 2024.

\bibitem[Trautmann et~al.(2016)Trautmann, Rehm, and Wittchen]{Trautmann_2016}
S.~Trautmann, J.~Rehm, and H.~Wittchen.
\newblock The economic costs of mental disorders.
\newblock \emph{The British Journal of Psychiatry}, 17:\penalty0 1245–1249, September 2016.

\bibitem[Wang et~al.(2024)Wang, Ouyang, R.~Jiao, Zhang, Shang, Jia, Yan, Wu, and Liu]{Wang_2024}
J.~Wang, H.~Ouyang, S.~Cheng R.~Jiao, H.~Zhang, Z.~Shang, Y.~Jia, W.~Yan, L.~Wu, and W.~Liu.
\newblock The application of machine learning techniques in posttraumatic stress disorder: a systematic review and meta-analysis.
\newblock \emph{npj Digital Medicine}, 7:\penalty0 121, May 2024.

\bibitem[Wright-Berryman et~al.(2023)Wright-Berryman, Cohen, Haq, Black, and Pease]{Wright-Berryman_2023}
J.~Wright-Berryman, J.~Cohen, A.~Haq, D.~P. Black, and J.~L. Pease.
\newblock The use of artificial intelligence for delivery of essential health services across who regions: a scoping review.
\newblock \emph{Frontiers in Psychiatry}, 14:\penalty0 1143175, June 2023.

\bibitem[Y.~Liu(2024)]{Liu_2024}
P.~A.~Bath Y.~Liu, S.~Mazumdar.
\newblock An unsupervised learning approach to diagnosing alzheimer’s disease using brain magnetic resonance imaging scans.
\newblock \emph{International Journal of Medical Informatics}, 173:\penalty0 105027, May 2024.

\bibitem[Yan et~al.(2023)Yan, Ruan, and Jiang]{Yan_2023}
W.~Yan, Q.~Ruan, and K.~Jiang.
\newblock Challenges for artificial intelligence in recognizing mental disorders.
\newblock \emph{Diagnostics}, 13:\penalty0 2, January 2023.

\end{thebibliography}

\end{document}